\documentclass[prl,superscriptaddress,showpacs,preprint,unsortedaddress]{revtex4}
\usepackage{graphicx}
\usepackage{dcolumn}
\usepackage{amsmath}
\usepackage{hyperref}
\usepackage{bm}
\usepackage{color}
\usepackage{ulem}

\newcommand{\chem}[1]{\ensuremath{\mathrm{#1}}}
\newcommand{\un}[1]{\ensuremath{\,\mathrm{#1}}}

\newcommand{\TbVO}[0]{\chem{TbVO_4}}

\renewcommand{\vec}[1]{\ensuremath{\bm{#1}}}

\begin{document}

\title{Direct observation of the high magnetic field effect on the Jahn-Teller state in \TbVO}

\author{C. Detlefs}

\affiliation{European Synchrotron Radiation Facility, B.P. 220, F--38043 Grenoble Cedex, France}

\author{F. Duc}

\affiliation{Laboratoire National des Champs Magn\'etiques Puls\'es,
  143, avenue de Rangueil, F--31400 Toulouse, France}

\author{Z.~A. Kaze{\u\i}}

\affiliation{Moscow State University, Leninskye Gory, Moscow, 119992 Russia}

\author{J. Vanacken}

\affiliation{Pulsveldengroep, Institute for Nanoscale Physics and
  Chemistry, Clestijnenlaan 200D, B--3001 Leuven, Belgium}

\author{P. Frings}

\affiliation{Laboratoire National des Champs Magn\'etiques Puls\'es,
  143, avenue de Rangueil, F--31400 Toulouse, France}

\author{W. Bras}

\affiliation{Netherlands Organisation for Scientific Research (NWO),
  DUBBLE CRG at the ESRF, BP220, F--38043 Grenoble, France}

\author{J.~E. Lorenzo}

\affiliation{Institut N\'{e}el, CNRS, BP 166X, F--38043
  Grenoble, France}
  
\author{P.~C. Canfield}

\affiliation{Ames Laboratory, USDOE and Department of Physics and
Astronomy, Iowa State University, Ames, IA 50011, USA}

\author{G.~L.~J.~A. Rikken}

\affiliation{Laboratoire National des Champs Magn\'etiques Puls\'es,
  143, avenue de Rangueil, F--31400 Toulouse, France}

\date{Version \today}


\begin{abstract}
  We report the first direct observation of the influence of  high magnetic 
  fields on the Jahn-Teller (JT) transition in \TbVO{}. Contrary to 
  spectroscopic and magnetic methods, X-ray diffraction directly measures 
  the JT distortion; the splitting between the $(311)/(131)$ and 
  $(202)/(022)$ pairs of Bragg reflections is proportional to the order 
  parameter. Our experimental results are compared to mean field 
  calculations, taking into account all possible orientations of the 
  grains relative to the applied field, and qualitative agreement is obtained.
\end{abstract}

\maketitle

%\section{Introduction}

Phase transitions induced by quadrupolar interactions have recently received much 
attention, mostly due to their influence on the magnetic properties of the sample, 
e.g.~in the giant-magneto-resistive effect of manganite compounds \cite{Dagotto03} 
and the magneto-electric effect in multi-ferroic materials \cite{Hur04,Chapon04}. 
The cooperative Jahn-Teller (JT) effect arises from the same quadrupolar interactions 
\cite{Gehring75}, and may indeed be interpreted as ferro-quadrupolar order. In JT 
compounds, the balance between magnetic and quadrupolar effects can be tuned by varying 
the strength of an externally applied magnetic field. To date, however, only a small amount
of experimental evidence has been reported \cite{Mangum71}, as the required magnetic fields tend 
to be rather large.

In this letter we present the first direct experimental evidence for the modification of the JT 
effect by a external magnetic field.  This effect was predicted by both qualitative 
\cite{Sivardiere72,Sivardiere72b} and quantitative \cite{Demidov05,Kazei05} theories, 
but previous experiments \cite{Mangum71,Pytte74,Hartley77,Kutko96,Kazei05} observed only indirect signatures 
of this behavior.

Terbium ortho-vanadate, \chem{TbVO_4}, along with \chem{DyVO_4}, is a textbook example 
for a JT transition \cite{Gehring75} induced by quadrupolar interactions between the 
\chem{Tb} $4f$ moments, mediated through phonons.
At high temperatures, \chem{TbVO_4} crystallizes in the tetragonal zircon structure with 
space group $I4_1/amd$ \cite{Kirschbaum99}, with lattice parameters $a_t=b_t=7.1831(3)\,${\AA}
and $c=6.3310(4)\,${\AA}. Upon lowering the temperature through $T_{Q} \approx 33\un{K}$, 
it undergoes a cooperative JT transition: The crystal spontaneously distorts along the $[110]$ 
direction ($B_{2g}$ strain, $\delta$-symmetry distortion) to the orthorhombic space group 
$Fddd$ with lattice parameters $a_o=10.239(2)\,${\AA}, $b_o=10.029(2)\,${\AA}, and 
$c=6.3154(13)\,${\AA} at $22\un{K}$. The distortion is surprisingly large, reaching 
$\epsilon^{\delta} = 2(a_o-b_o)/(a_o+b_o)=2\%$ at $22\un{K}$ \cite{Kirschbaum99}, and increasing 
further toward lower temperatures. 

The JT transition has been studied extensively, both experimentally and theoretically. 
The early studies, however, focused on the sample's properties in the absence of applied 
magnetic fields \cite{Gehring75}. 
The effect of large external magnetic fields on {\TbVO} was studied only recently.  Quantitative 
mean-field calculations \cite{Demidov05} found that the JT distortion is suppressed when fields 
above $\approx 29\un{T}$ are applied along the $c$-axis of the sample, in good agreement with 
susceptibility measurements \cite{Kazei05}. Diffraction experiments have so far not been carried 
out, as the required combination of high magnetic fields and X-ray diffraction equipment has only 
recently become available \cite{Matsuda04,Narumi06,Frings06}.

The experiments were performed on the DUBBLE CRG beamline, BM26B, at the ESRF \cite{Bras03}, using 
the experimental configuration described in \cite{Frings06}. Flux-grown single crystals 
of \chem{TbVO_4} \cite{Smith74,Wanklyn78} were ground into a fine powder and embedded in low 
molecular weight polyvinylpyrolidone in order to prevent movement of the powder grains due to 
magnetic forces, while at the same time improving the thermal contact. The sample was then mounted 
in a pulsed magnet and cryostat assembly \cite{Frings06}.

\begin{figure}
    \includegraphics[width=1\columnwidth]{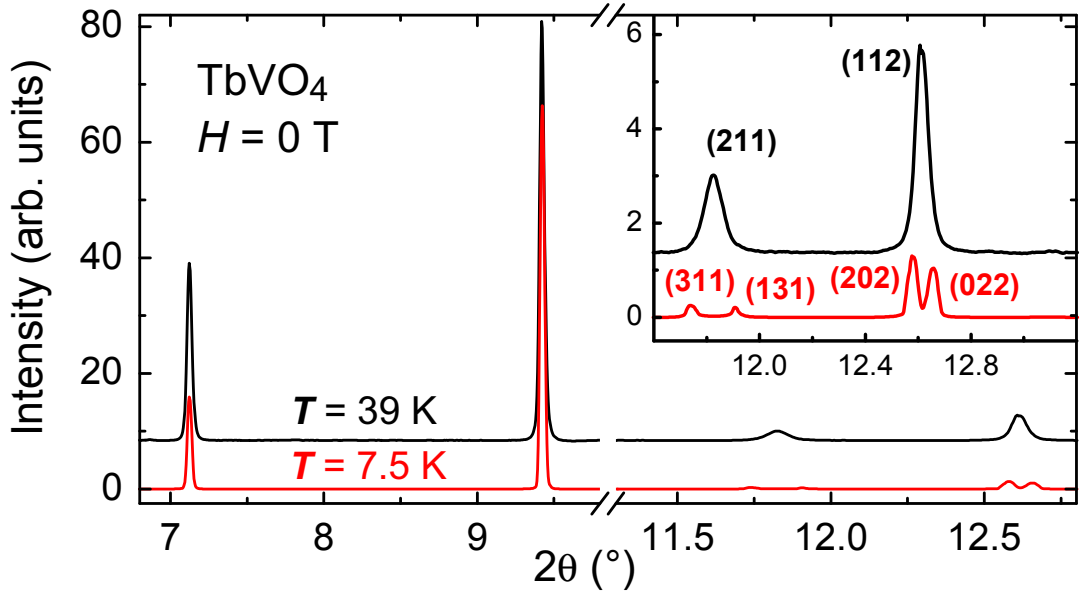}
	\caption[X-ray powder diffraction spectra of {\TbVO}]{\label{fig.spectra}X-ray powder diffraction spectra 
		of {\TbVO}. The inset shows the $(311)/(131)$ and $(202)/(022)$ pairs of reflections that are sensitive 
		to the orthorhombic distortion, $\epsilon^\delta$.
	}
\end{figure}

The data shown here were taken by accumulating $\approx 45$ magnetic field pulses per spectrum. 
For each field pulse a mechanical shutter exposed the image plate detector for $4.9\un{ms}$ centered 
around the maximum field. Representative powder diffraction spectra are shown in Fig.~{\ref{fig.spectra}}.

The JT transition manifests itself as splitting of some, 
but not all, powder lines (see table~\ref{tab.peaks}). For small distortions, $\epsilon^{\delta} \ll 1$, 
the orthorhombic lattice parameters can be approximated as 
$a_{\mathrm{o}} \approx (1+\frac{1}{2}\epsilon^{\delta})\bar{a}$ and 
$b_{\mathrm{o}} \approx (1-\frac{1}{2}\epsilon^{\delta})\bar{a}$, where 
$\bar{a}=\frac{1}{2}(a_{\mathrm{o}} + b_{\mathrm{o}}) \approx
\sqrt{2}a_{\mathrm{t}}$. To first order in $\epsilon^{\delta}$, the splitting of a pair of Bragg 
reflections $(H,K,L)_{\mathrm{o}}$ and $(K,H,L)_{\mathrm{o}}$ is then given by
\begin{equation}
  \delta(2\theta)
  \approx 
  \frac{\lambda^2}{\bar{a}^2}
  \frac{ K^2-H^2 }{\sin(2\theta)}
  \epsilon^{\delta} 
  =  \tan(\theta) \frac{K^2-H^2}{H^2+K^2+\frac{\bar{a}^2}{2{c}^2}L^2}
  \epsilon^{\delta}.
  \label{eq.epsilon_tth}
\end{equation}

\begin{table}
  \caption[Bragg reflections]{\label{tab.peaks}Bragg reflections of
    {\TbVO}, scattering angle at $E=21\un{keV}$, and the splitting due
    to the JT-effect. Values are based in the lattice parameters given
    in Ref.~\onlinecite{Kirschbaum99} for $T=22\un{K}$.}
  \begin{tabular}{c|c|c}
    $(HKL)_o$ & $2\theta               $ & $\delta(2\theta) / \epsilon^{\delta}$ \\
            & $        [\mathrm{deg}]$ & $  [\mathrm{deg}/\%]$ \\
    \hline
    $(111)$       & 7.14 & 0 \\
    $(220)$       & 9.45 & 0 \\
    $(311)/(131)$ & 11.78/11.94 & 0.076  \\
    $(202)/(022)$ & 12.61/12.69 & 0.036  \\
    $(400)/(040)$ & 13.24/13.52 & 0.134  \\
    $(222)$       & 14.32 & 0 \\
    $(331)$       & 15.19 & 0 \\
    $(113)$       & 16.81 & 0 \\
  \end{tabular}
\end{table}

Our experimental configuration, with photon energy $E=21\un{keV}$ ($\lambda=0.59\,${\AA}), allowed us to observe scattering 
angles up to $2\theta=12.8^\circ$, covering the reflections, in the orthorhombic notation, $(111)$, $(220)$, 
$(311)$, $(131)$, $(202)$, and $(022)$. Within this range, only the $(311)/(131)$ and the $(202)/(022)$ 
pairs are sensitive to the JT distortion.

In samples with large magneto-crystalline (MC) anisotropy, such as {\TbVO}, the phase diagram depends not only 
on the magnitude of an applied magnetic field, but also on its orientation relative to the crystal symmetry axes. 
The grains in a powder sample are aligned randomly, so that for a macroscopic measurement, 
e.g.~magnetic susceptibility, the direction of the field has to be averaged over the full solid angle of $4\pi$.

In a diffraction experiment, however, the orientation is constrained: Any grain contributing to the diffraction 
signal at the angle $2\theta$ must be oriented such that the Bragg planes of the corresponding reflection 
$(H,K,L)$ form the angle $\theta$ with the incident beam. The powder averaging must then be taken only over 
the azimuthal angle, $\psi$, i.e. rotation about the scattering vector $\vec{Q}=(H,K,L)$.

In our experiment the external field, $\mu_0 \vec{H}$, was aligned along the incident beam axis; the angle 
between the field and the Bragg planes is therefore the Bragg angle, $\theta$. 

For systems with substantial MC anisotropy any simulation of the powder diffraction 
spectrum must thus be carried out peak by peak: For each 
Bragg reflection the magnetic field direction and the corresponding influence on the phase diagram and 
the lattice parameters must be calculated as function of $\psi$, yielding, in this case the distortion 
$\epsilon^\delta(\psi)$, and the corresponding scattering angle, $2\theta_{\mathrm{calc}}(HKL, \epsilon^\delta(\psi))$. 
The resulting diffraction patterns are then averaged over the azimuthal angle, taking into account the instrumental resolution.

In order to quantitatively describe the effect outlined above for {\TbVO} we have performed comprehensive mean field 
calculations of the magnetoelastic distortion of {\TbVO} as function of the strength and direction 
of the externally applied magnetic field. Our calculations closely follow those of \cite{Demidov05} 
and \cite{Kazei05} and will be published in full detail elsewhere. In particular, we have used 
crystal field parameter set 2 of \cite{Demidov05}, and a quadrupolar constant $G^\delta = 130\un{mK}$.

The order parameter (OP), $\epsilon^{\delta}$, is proportional to the expectation value of the quadrupole 
operator \footnote{Note that for consistency with the existing literature \cite{Demidov05,Kazei05} we have 
	defined the quadrupole operator $P_{xy}$ in the high-temperature, tetragonal coordinate system. In 
	the low-temperature, orthorhombic basis $\epsilon^\delta \propto \langle O_2^2\rangle$, with 
	$O_2^2 = J_x^2-J_y^2$.
}, $Q_{xy} = \langle P_{xy} \rangle$, 
\begin{equation}
	\epsilon^\delta = \frac{B^\delta}{C^\delta_0} Q_{xy},
	\label{eq.op}
\end{equation}
where $P_{xy} = \frac{1}{2}(J_x J_y + J_y J_x)$. The coefficient 
$B^\delta/C^\delta_0 = 13.9 \cdot 10^{-4}$ was determined from the zero-field, low temperature value 
of the spontaneous deformation, $\epsilon^\delta$.

We first consider the influence of the applied magnetic field on the quadrupole moment $Q_{xy}$. 
Two effects must be taken into account: Changes in the magnitude of the distortion will result in 
variations of the splitting between the pairs of peaks, whereas (for a given grain) the preferential 
population of domains with positive or negative distortion will lead to a magnetically induced texture 
that manifests itself in a shift of spectral weight from one partner to the other.

A magnetic field along the $c$ axis leads to a non-vanishing expectation value of $\langle J_z \rangle$ in 
competition with the OP. A magnetic field along $a_{\mathrm{o}}$ or $b_{\mathrm{o}}$, on the other 
hand, will induce a magnetization $\langle J_x \pm J_y \rangle$ along the corresponding axis. Above the JT 
transition a magnetostrictive orthorhombic distortion of {$\delta$}-symmetry is induced. 
In the ordered state, $T < T_Q$, however, one has to distinguish between a field applied along 
$a_{\mathrm{o}}$ and $b_{\mathrm{o}}$: Along the two axes, the susceptibility will be of slightly different value, 
thus breaking the degeneracy of $\pm \epsilon^{\delta}$. Crossing the phase boundary in the presence of such 
a magnetic field will then lead to the preferential population of one of these domains. Fields along the 
$[110]_{\mathrm{o}}$ directions should lead to a distortion with {$\gamma $}-symmetry 
(B$_{1g}$-type), as observed in \chem{DyVO_4} \cite{Kazei98,Demidov04}. Finally, if the external 
field is applied along an arbitrary direction, then both $\delta$ and $\gamma$ strains are induced 
simultaneously, lowering the symmetry of the crystal to monoclinic. Fortunately, the $\gamma$ 
magnetostrictive distortion in {\TbVO} is two orders of magnitude smaller than the $\delta$ distortion and can 
therefore be neglected.

\begin{figure}
		\includegraphics[width=1\columnwidth]{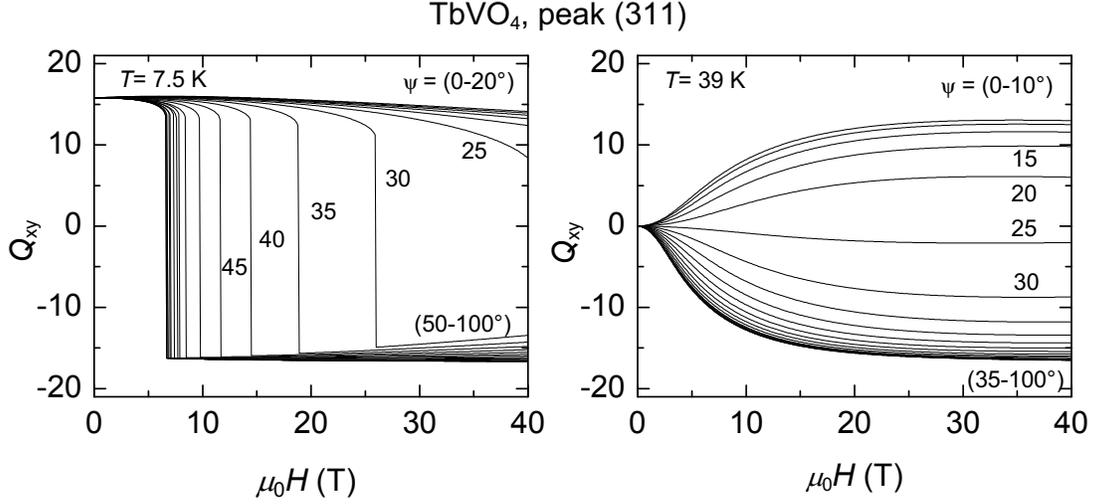}
	\caption[Calculated dependence of the quadrupole moment on the external magnetic field for different azimuthal angles.]{
		\label{fig.qxy.311}Calculated dependence of the quadrupole moment $Q_{xy}$ on the external magnetic field for 
			different azimuthal angles in the interval (0--100$^\circ$) at the diffraction peak $(311)$. Left: $T=7.5\un{K}$, Right: $T=39\un{K}$.
		}
\end{figure}

Typical theoretical results for the $(311)$ diffraction peak at 
$T=7.5\un{K}$ and $39\un{K}$ are shown in Fig.~\ref{fig.qxy.311}. The $(311)$ partner of 
the $(311)$/$(131)$ pair corresponds to a positive quadrupole moment, $Q_{xy} > 0$ (see~eq.~\ref{eq.epsilon_tth}). 

At $T=39\un{K} > T_Q$ there is no spontaneous quadrupole moment at $H=0$. For azimuthal angles 
$\psi$ below $\approx 22^\circ$ a positive quadrupole moment is induced, whereas for larger 
azimuthal angles the induced quadrupole moment is negative, i.e.~corresponding to the $(131)$ 
partner. We thus expect the magnetic field to induce a magnetoelastic splitting with unequal 
domain populations of $\approx 33\%$ $Q_{xy}>0$ (contributing to the $(311)$ diffraction peak) 
and $\approx 67\%$ $Q_{xy}<0$ (contributing to the $(131)$ diffraction peak) when averaging over the full interval (0--360$^\circ$).

At $T=7.5\un{K} < T_Q$ the situation is different: A sizable spontaneous quadrupole moment is 
present, and in a zero-field cooled sample domains with positive and negative $Q_{xy}$ will be 
populated almost equally. Upon applying an external field the quadrupole moment is reduced, 
and the degeneracy of $\pm Q_{xy}$ is lifted, so that (depending on their azimuthal angle) domains 
may invert their quadrupole moment. This transition is first order and therefore 
subject to pinning on defects, grain boundaries, surfaces, and elastic interactions between twinned 
domains \cite{Kirschbaum99}. 

\begin{figure}
		\includegraphics[width=1\columnwidth]{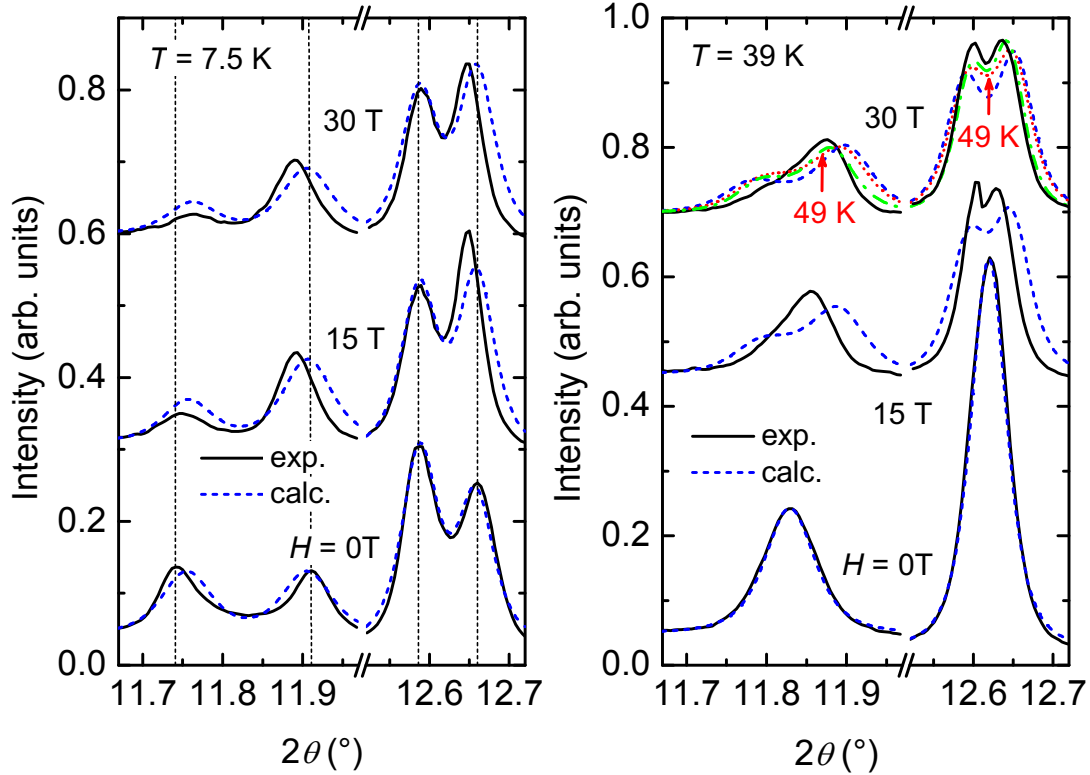}
	\caption[Comparison between calculated and measured spectra for different temperatures and fields]{
		\label{fig.compare}Comparison between calculated and measured spectra for different temperatures 
		and fields. Left: $T=7.5\un{K}$. Right: $T=39\un{K}$. At $T=39\un{K}$ and $30\un{T}$, the dotted curve 
		(in red) corresponds to the calculations including the magnetocaloric effect, and the green line to the 
		calculations with the coefficient $B^{\delta } / C_{0}^{\delta }$ reduced by $25{\%}$.
	}
\end{figure}

\begin{figure}
		\includegraphics[width=1\columnwidth]{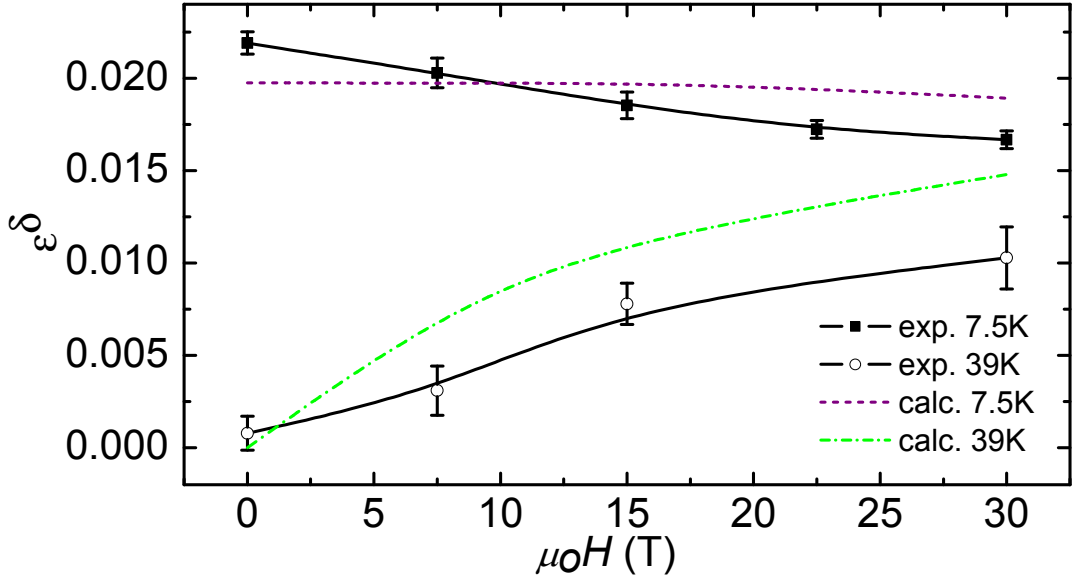}
	\caption[Average order parameter as function of applied magnetic field for selected temperatures]{
		\label{fig.avg_eps}Average order parameter as function of applied magnetic field for selected 
		temperatures. Continuous lines: Experimental data. Dashed lines: Theory
	} 
\end{figure}

The calculated dependences $Q_{xy}(H)$ for fixed $T$ and different azimuthal angles $\psi$ allow us to simulate 
the X-ray powder diffraction data. For the $(202)/(022)$ pair the values were 
averaged over the interval $0^\circ < \psi < 180^\circ$ since the two-fold symmetry persists. For the 
$(311)/(131)$ pair the averaging was performed over $0^\circ < \psi < 360^\circ$. The parameters of the 
resolution function were fitted to the experimental data 
at $T>T_Q$ and $H=0$. Different parameter sets were used for the $(311)/(131)$ and $(202)/(022)$ pairs, 
and the intensity parameter had to be adjusted to match the data at low temperature and high field.

At $T > T_{Q}$, calculations for the $(202)/(022)$ pair adequately reproduce the almost symmetric 
splitting of the line by the magnetic field observed in the data (Fig.~\ref{fig.compare}). The calculated
magnitude of the splitting, however, is $\approx 25{\%}$ larger than the experimental one 
(Fig.~\ref{fig.avg_eps}). The origin of this discrepancy remains unclear. In particular, it does not seem 
to be related to the value of the interaction parameters used in the calculations since the value of the 
spontaneous splitting at $H = 0$ is correctly described (Fig.~\ref{fig.avg_eps}). Similar deviations 
between theory and experiment are observed at the $(311)/(131)$ pair. Calculations describe the shape of 
the $(311)$ line in different fields: the intensity ratio for two components of the split peak is about 1:2. 
But again the calculated value of the splitting is larger than the experimental one by about $25{\%}$ (see Fig.~\ref{fig.compare}). The OP may be decreased by sample heating, e.g.~due to the 
magneto-caloric effect. A reasonable temperature increase of about $8$--$10\un{K}$, however, cannot fully 
explain the observed discrepancy. 

The experimental observations at $T > T_Q$ and $H = 30\un{T}$ are well described by our calculations when the 
coefficient $B^{\delta } / C_{0}^{\delta }$ is reduced by $25{\%}$; this value, however, is too small for the 
spontaneous deformation. This may be interpreted as the variation of the quadrupolar constant $G^{\delta }$ 
under high magnetic field, as was observed for the quadrupolar constant $G^{\gamma }$ of \chem{DyVO_4} 
and considered within an improved "compressible Ising model" \cite{Demidov04,Page79}. However for 
\chem{TbVO_4} no dependence of $G^{\delta}$ was observed when the OP increases as the temperature decreases, 
and its low field behavior is described adequately by the mean field theory \cite{Gehring75}. 

For the quadrupole ordered phase the results of the numerical calculations for the $(202)$ and $(311)$ peaks 
 are compared with the experimental data in Fig.~\ref{fig.compare}. 
For both peaks the averaging were performed for two domains ($\epsilon^{\delta}$ and $-\epsilon^{\delta }$) 
for $0 < \psi < 360^\circ$, in steps of $\Delta\psi=5^\circ$. For the (202)/(022) pair the calculations do not reveal a noticeable decrease of the splitting 
under external field of $30\un{T}$, while for the $(311)/(131)$ pair a small shift is expected for the low-angle 
component.

We examined, and refuted, several possible explanations for the discrepancy between the theoretical and experimental results.
Firstly we ascertained that various weak pair interactions (bilinear, quadrupole of ${\gamma}$-, 
${\alpha}$- and ${\epsilon}$- symmetry) which were omitted in our calculations do not change the results 
noticeably. Then we estimated the effect of the sample heating due to the magnetocaloric effect in our 
experiment. In the adiabatic regime, the temperature change is large enough (about $25\un{K}$ at $7.5\un{K}$) for the magnetic field orientation close to the easy magnetization axis. Additional studies are 
necessary to elucidate the influence of the magnetocaloric effect in our experiments. 

In summary, we have directly observed for the first time the effect of magnetic fields on the Jahn-Teller distortion of {\TbVO}, and compared the experimental data to a detailed mean field theory. Due to the polycrystalline nature of the sample and its strong MC anisotropy the calculated spectra had to be averaged over all possible orientations of the powder grains relative to the applied magnetic field. The applied magnetic field was found to influence both the magnitude of the OP, as observed in the splitting of the $(311)/(131)$ and $(202)/(022)$ pairs of Bragg peaks, and the relative domain populations, reflected in the intensity ratio between the partners of a pair. Our theory is in qualitative agreement with the experimental results, even though small quantitative discrepancies persist.

This new technique was found to be a powerful tool to study JT systems, and allows us to observe new aspects of magnetic behavior of the classical and well studied JT compound {\TbVO}. These high magnetic field data may be used both to revise known theoretical models and to develop new, improved ones.

\acknowledgments

The authors acknowledge the NWO/FWO Vlaanderen and ESRF for granting
the beamtime for these experiments, and thank the staff of DUBBLE CRG
and the ESRF for help in setting up these experiments. 
We thank M.~Nardone, J.~Billette, and A.~Zitouni for their
excellent work on the pulsed field coil and cryostat.
Ames Laboratory is supported by the Department of Energy,
Office of Science under Contract No.~W-7405-Eng-82. Z.~A.~K.~acknowledges 
financial support from the RFBR (project No.~07-02-01043). Part of this research was 
funded by the ANR (ANR-05-BLAN-0238) and EuroMagNet (EU contract 506239).

%\bibliography{../../../allref}

\bibliography{references}

\end{document}